# Dynamic physical layer equalization in optical communication networks

TIANHUA XU[a,*], GUNNAR JACOBSEN[b], JIE LI[b], MARK LEESON[a], SERGEI POPOV[c]
*aUniversity of Warwick, Coventry, CV4 7AL, United Kingdom*
*bRISE Acreo AB, Stockholm, SE-16440, Sweden*
*cRoyal Institute of Technology, Stockholm, SE-16440, Sweden*

In optical transport networks, signal lightpaths between two terminal nodes can be different due to current network conditions. Thus the transmission distance and accumulated dispersion in the lightpath cannot be predicted. Therefore, the adaptive compensation of dynamic dispersion is necessary in such networks to enable flexible routing and switching. In this paper, we present a detailed analysis on the adaptive dispersion compensation using the least-mean-square (LMS) algorithm in coherent optical communication networks. It is found that the variable-step-size LMS equalizer can achieve the same performance with a lower complexity, compared to the traditional LMS algorithm.

## 1. Introduction

The performance of high speed optical fiber networks is significantly affected by system impairments from chromatic dispersion (CD), polarization mode dispersion (PMD), laser phase noise, and fiber nonlinearities [1-14]. Due to the high transmission spectral efficiency and the robust tolerance to fiber nonlinearities, coherent optical detection employing advanced modulation formats and digital signal processing (DSP) has become one of the most promising solutions for the next generation of high speed optical fiber networks [15-20]. Since both the amplitude and the phase information from the received signal can be extracted using coherent optical detection, transmission impairments such as those above can be compensated or mitigated effectively using powerful DSP algorithms [21-33]. Chromatic dispersion can be well compensated and equalized using time-domain and frequency-domain digital filters [21-24], which have become the most promising alternative approaches to dispersion compensating fibers (DCFs) [1,2]. These implementations lead to a dramatic reduction in the complexity and costs, as well as increased tolerance to fiber nonlinearities, for high-capacity optical fiber transmission networks.

To date, a number of digital equalizers have been implemented based on a fixed amount of fiber dispersion to realize static compensation of inter-symbol interference (ISI), where an accurate knowledge of chromatic dispersion in the transmission link is critically required [21-24]. However, in switched optical fiber networks, the signal lightpath between two terminal nodes can change over time according to different network conditions, where the transmission distance and the accumulated dispersion in the lightpath cannot be predicted in advance. Therefore, adaptive compensation for the chromatic dispersion in such optical transmission networks should be given serious consideration. Recently, adaptive CD equalization in dynamically switched optical networks has attracted some research interest, and several approaches such as the least-mean-square (LMS) algorithm, the constant modulus algorithm (CMA), the delay tap sampling technique, the overlap frequency domain equalization, and the auto-correlation of signal power waveform have been investigated to enable a flexible routing and switching in such optical fiber networks [34-38]. Among these methods, the time-domain LMS equalizer can deliver a relatively simple specification and a large dynamic range, as well as a good tolerance to small laser phase noise, and thus becomes a very promising solution for the adaptive CD electronic equalization in dynamically switched and routed optical fiber networks [37-40].

In this paper, we present a detailed analysis of adaptive chromatic dispersion compensation using the LMS algorithm, in coherent optical fiber transmission networks. Numerical simulations have been carried out in the dual-polarization quadrature phase shift keying (DP-QPSK) coherent transmission system, based on the VPI and Matlab platforms [41,42]. The influence of step size in the LMS adaptive equalization for compensating the chromatic dispersion is investigated, and the impact of step size on the tap weight convergence in the LMS equalizer is also analyzed in detail. The LMS filter shows a better CD compensation performance by using a smaller step size, but this will result in a slower iterative computation to achieve the convergence of the tap weights. To solve this contradiction, a variable-step-size LMS (VSS-LMS) algorithm is further proposed to realize the dynamic equalization of the chromatic dispersion in coherent optical fiber networks. The performance of the VSS-LMS for adaptive CD compensation is analyzed and compared to the traditional LMS filter, and the required number of taps and the distribution of converged tap weights in both equalizers for a specific fiber dispersion profile are also investigated. It is found that the VSS-LMS adaptive equalizer can give an optimum CD equalization performance compared to the traditional LMS algorithm, where a good compromise between the CD compensation performance and the converging speed of the tap weights can be obtained. Therefore, the VSS-LMS equalization can achieve an optimum balance between the CD equalization performance and the computational complexity, where the best CD compensation with a low complexity can be realized.

## 2. Principle of least-mean-square based adaptive dispersion compensation

In this section, the principle of the traditional LMS algorithm and the variable-step-size LMS algorithm are described, and the influence of step size on the update and the convergence of the tap weights are also discussed in detail.

### 2.1. Structure of least-mean-square based adaptive equalizers

The schematic of the adaptive equalizer based on the LMS algorithm with a number of tap weights, $N$, is illustrated in Fig. 1, where $T$ is the sampling period, $W_i$ ($i$=1,2,…,$N$) represents the tap weight coefficient in the LMS based equalizer, $x_i$ is the input sample sequence, $y$ is the equalized output sample, $d$ is the desired output sample, and $e$ is the estimation error between the output $y$ and the desired output $d$.

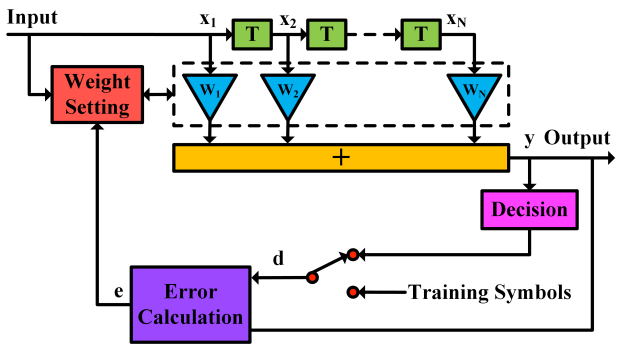

*Fig. 1. Block diagram of the LMS algorithm based adaptive equalizer for dispersion compensation*

As shown in Fig. 1, the adaptive equalizer includes a tapped delay line for storing the data samples from the input signal sequence. During each sample period, the adaptive equalizer calculates the convolution between the tap weights in the delay line and the input samples, and then the tap weights are updated for the calculation in the next sample period. The tap weights are updated according to the estimation error between the output signal and the desired signal, and the speed of the update depends on the step size parameter. Meanwhile, the adaptive equalizer can be applied in the decision-direct (DD) or the training symbol modes; here, the DD, update option is employed in our analysis and numerical simulations.

### 2.2. Principle of LMS adaptive algorithm

The LMS equalizer is a branch of the family of adaptive algorithms, which is designed by finding the filter coefficients to produce the least mean square value of the error signal (the difference between the desired output and the actual output signal). The LMS algorithm is an iterative adaptive method which can be applied in highly time-varying signal environments. It is a stochastic gradient descent approach, since the tap weights in the LMS filter are only accommodated based on the current estimation error. The traditional LMS algorithm incorporates an iterative procedure which makes successive corrections to the tap weight vector in the negative direction of the gradient vector which eventually results in a minimum mean squared error. The equalized output signal and the tap weights vector of the LMS adaptive equalizer can be expressed as follows [38-40],

$$y(n) = \vec{w}_{LMS}^{H}(n)\vec{x}(n) \tag{1}$$

$$\vec{w}_{LMS}(n+1) = \vec{w}_{LMS}(n) + \mu_{LMS}\,\vec{x}(n)e_{LMS}^{*}(n) \tag{2}$$

$$e_{LMS}(n) = d_{LMS}(n) - y(n) \tag{3}$$

where $\vec{x}(n)$ is the vector of the complex input signal, $y(n)$ is the equalized complex output signal, $\vec{w}_{LMS}(n)$ is the vector of the complex tap weights, $d(n)$ is the desired output symbol, $e(n)$ is the estimation error between the output signal $y(n)$ and the desired symbol $d(n)$, $\mu_{LMS}$ is the step size parameter which controls the convergence characteristics of the LMS algorithm, $H$ represents the Hermitian transform, and * means the conjugate operation. The tap weight vector $\vec{w}(n)$ is firstly initiated with an arbitrary value $\vec{w}(0)$ at $n$=0, and then is updated in a sample-by-sample (or symbol-by-symbol) iterative manner to achieve the eventual convergence, when the estimation error $e(n)$ approaches zero.

In order to guarantee the convergence of $\vec{w}(n)$ in the LMS equalizer, the step size parameter $\mu_{LMS}$ in the adaptive filter needs to satisfy a condition of $0 < \mu_{LMS} < 1/\lambda_{max}$, where $\lambda_{max}$ is the largest eigenvalue of the correlation matrix $R = \vec{x}(n)\vec{x}^{H}(n)$ [38-40]. The convergence speed of the algorithm is inversely proportional to the eigenvalue spread of the correlation matrix $R$. The convergence of the LMS tap weights will be slow, when the eigenvalues are widely spread. The eigenvalue spread of the correlation matrix $R$ is evaluated by calculating the ratio between the largest eigenvalue and the smallest eigenvalue. The LMS algorithm will converge quite slowly, when the step size $\mu_{LMS}$ is very small. One the other hand, the LMS algorithm will converge faster for a larger value of step size $\mu_{LMS}$. Howerver, the LMS algorithm can be less stable since sometimes the step size may exceed $1/\lambda_{max}$.

## 2.3. Principle of variable-step-size LMS adaptive algorithm

Generally, the traditional LMS algorithm is quite robust for dispersion compensation and requires only a small computational effort. However, the accommodation of the step size will impact both the convergence speed and the residual error in the traditional LMS equalizer. The performance of the traditional LMS algorithm can be enhanced and optimized, if the step size of this adaptive equalizer can be adjusted properly. For the best situation, a larger step size is applied at the beginning of the process to accelerate the convergence speed, and a smaller step size is applied after approximate convergence to generate the smallest residual error. Correspondingly, the variable-step-size LMS algorithm has been developed to improve the performance of the traditional LMS algorithm in terms of the convergence speed and residual error level [39,40,43-45]. The step size parameter in the VSS-LMS algorithm changes with the variation of the mean square error, which allows the adaptive equalizer to track the changes in the transmission system as well as to produce a small steady residual error. The equalized output signal $y(n)$ and the tap weights vector of the variable-step-size LMS adaptive filter can be expressed as the following equations [39,40],

$$y(n) = \overset{\rightarrow}{w}{}^{H}_{VSS-LMS}(n)\overset{\rightarrow}{x}(n) \tag{4}$$

$$\overset{\rightarrow}{w}_{VSS-LMS}(n+1) = \overset{\rightarrow}{w}_{VSS-LMS}(n) \tag{5}$$
$$+ \mu_{VSS-LMS}(n)\overset{\rightarrow}{x}(n)e^{*}_{VSS-LMS}(n)$$

$$\mu_{VSS-LMS}(n+1) = \alpha\mu_{VSS-LMS}(n) + \gamma e^{2}_{VSS-LMS}(n) \tag{6}$$

$$e_{VSS-LMS}(n) = d_{VSS-LMS}(n) - y(n) \tag{7}$$

where $\overset{\rightarrow}{x}(n)$ is the vector of the complex input signal, $y(n)$ is the equalized output signal using the VSS-LMS filter, $\overset{\rightarrow}{w}_{VSS-LMS}(n)$ is the vector of the complex tap weights, $d(n)$ is the desired output symbol, $e(n)$ represents the estimation error between the output signal $y(n)$ and the desired symbol $d(n)$, and $\mu(n)$ is the step size coefficient of the VSS-LMS algorithm for adjusting the convergence properties and the residual error, and is updated with the variation of the estimated error $e(n)$. The parameters $\alpha$ and $\gamma$ are the coefficients for controlling the step size to be updated with the change of estimation error $e(n)$, and the range of the parameters are $0 < \alpha < 1$ and $\gamma > 0$  $\gamma > 0$. The convergence speed of the VSS-LMS adaptive algorithm can be accommodated by choosing different values for the energy attenuation factor $\alpha$.

The step size $\mu(n)$ is always positive and is controlled by the size of the estimated error and the parameters $\alpha$ and $\gamma$, according to Eq. (6). A typical value of $\alpha$=0.97 was found to work well in our numerical simulations, and the parameter $\gamma$ was usually chosen as $\gamma$=4.8×10$^{-4}$. In general, a large estimated error increases the step size to provide faster tracking. When the estimated error decreases, the step size will be decreased accordingly to reduce the misadjustment in estimation [38-40]. Compared to the traditional LMS algorithm, the VSS-LMS algorithm can give an improved performance at a cost of only four more multiplications or divisions in each iteration.

## 3. Implementation of DP-QPSK numerical transmission system

As illustrated in Fig. 2, a transmission arrangement comprising a 28-Gbaud DP-QPSK coherent optical communication system was numerically implemented using the VPI and Matlab platforms. All the simulations were carried out based on the nonlinear Schrödinger equation (NLSE) using the split-step Fourier solution. In the transmitter, the pseudo random bit sequence (PRBS) data from the 28-Gbit/s pattern generators were modulated into two orthogonally polarized QPSK optical signals by using Mach-Zehnder modulators and a polarization beam splitter (PBS). The orthogonally polarized signals were then fed into a standard single mode fiber (SSMF) transmission channel by using a polarization beam combiner to form the 28-Gbaud DP-QPSK optical signal. At the receiver end, the received optical signals were mixed with the local oscillator (LO) laser to be demodulated the baseband signals. The signals were detected by the photodiodes to become electrical signals and then digitalized by the analog-to-digital convertors (ADCs) at twice the symbol rate. Using DSP, system impairments in the transmission channel could be equalized and compensated using diverse digital filters. In this work, we neglected attenuation, polarization mode dispersion, laser phase noise, and fiber nonlinearities, since the investigation was only focused on the chromatic dispersion equalization. The bit error rate was evaluated based on $2^{18}$ bits, with a PRBS pattern length of $2^{15}$-1.

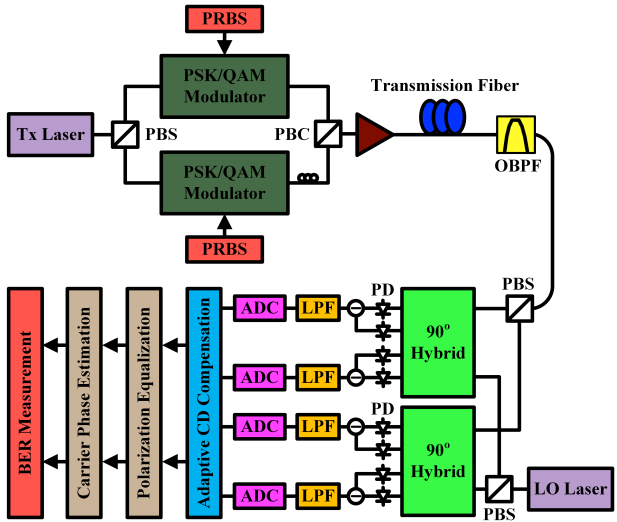

*Fig. 2. Schematic of the 28-Gbaud DP-QPSK coherent optical fiber transmission system. PBC: polarization beam combiner, OBPF: optical band-pass filter, LPF: low-pass filter*

## 4. Simulation results

To investigate the performance of VSS-LMS filter, the compensation of chromatic dispersion from a SSMF with a CD coefficient $D = 16$ ps/km/nm were numerically assessed, and the results compared to the traditional LMS adaptive filter. The tap weights were updated iteratively in both the traditional LMS algorithm and the variable-step-size LMS algorithm; here we mainly focused on the converged tap weights in the two equalizers. The converged tap weights of the LMS adaptive filter with 37 taps and step size of 0.1 for compensating the chromatic dispersion in the 60 km fiber are illustrated in Fig. 3. It can be seen that in the LMS adaptive filter, the central tap weights take more dominant roles in the chromatic dispersion equalization in all the tap weights magnitude, real part and imaginary part diagrams. For a fixed fiber dispersion, the tap weights in LMS adaptive filter approach zero, when the corresponding tap order exceeds a certain value, and this value indicates the least required taps number for compensating the chromatic dispersion effectively. This also illustrates the self-optimization characteristic of the least-mean-square adaptive algorithm. It could be seen from Fig. 3 that the required minimum number of taps in the LMS equalizer for equalizing 60 km fiber dispersion is 23.

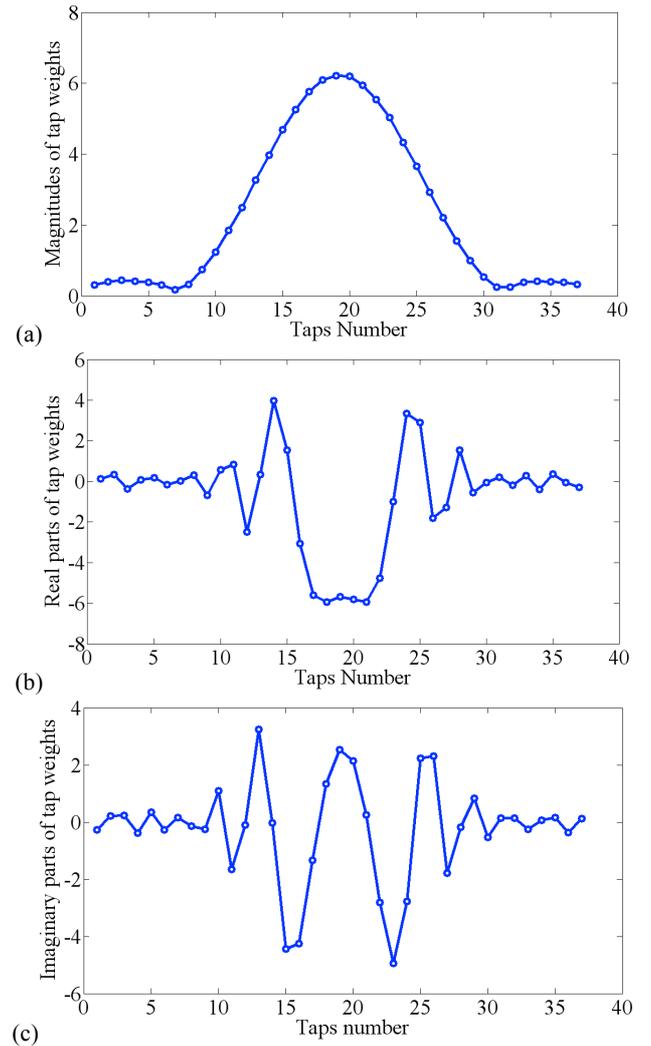

*Fig. 3. Tap weights of LMS adaptive filter (Tap orders are centralized) for 60 km fiber, (a) magnitudes of tap weights in LMS filter, (b) real parts of tap weights in LMS filter, (c) imaginary parts of tap weights in LMS filter*

The performance of CD compensation employing the LMS adaptive filter with step size value $\mu = 0.1$ using 9 taps for 20 km fiber dispersion and 2305 taps for 6000 km fiber dispersion is shown in Fig. 4. It could be seen from the figure that the two CD equalization results have little penalty compared with the back-to-back measurement when the fiber loss is neglected in the simulation work.

Simulation results of CD compensation employing LMS adaptive filter with different step size values using 401 taps for 1500 km fiber dispersion are shown in Fig. 5. It can be seen from this figure that the CD compensation results are better as the step size decreases, while a smaller step size will lead to a slower convergence speed. Also, it is observed that the BER performance is very similar, when the step size value is below $\mu = 0.1$, and the BER behavior become worse when the step size increases above $\mu = 0.1$. Therefore, the step size in the LMS adaptive equalizer was here usually selected as $\mu = 0.1$ to obtain the optimization.

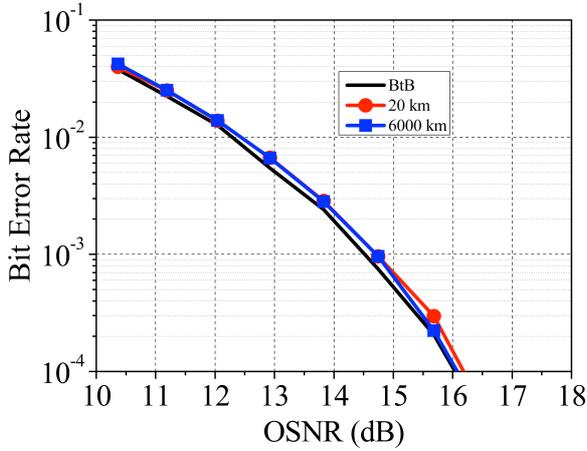

*Fig. 4. Chromatic dispersion compensation using LMS filter with a step size of 0.1 (neglecting fiber loss)*

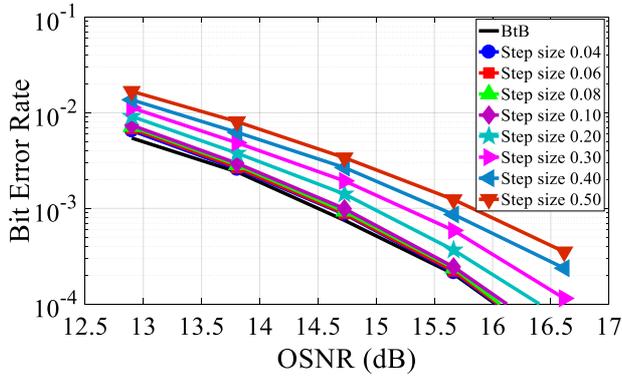

*Fig. 5. Chromatic dispersion compensation using LMS filter with different step sizes (neglecting fiber loss)*

The converged tap weights of the variable-step-size LMS adaptive filter for 60 km fiber dispersion with 37 taps and step size that varied between 0.06 and 0.6 are illustrated in Fig. 6. Again, it is seen that in the variable-step-size LMS adaptive filter, the central tap weights also take more dominant roles in the CD equalization in all the tap weights diagrams. It could also be found that the converged tap weights in the variable-step-size LMS filter vary consistently with the LMS adaptive filter tap weights, whereas the tap weights magnitudes in the variable-step-size LMS equalizer are larger than the tap weights magnitudes in the LMS equalizer.

To optimize the convergence speed and the compensation effect, the variable-step-size LMS algorithm was introduced and employed in the adaptive filter. The performance of the CD compensation for 60 km fiber dispersion using the variable-step-size LMS equalizer compared with the LMS equalizer is illustrated in Fig. 7. The VLMS adaptive equalizer could achieve the same CD compensation performance with the LMS adaptive equalizer, meanwhile, the VLMS filter uses step sizes varying from 0.06 to 0.6 that accelerates the algorithm converging speed.

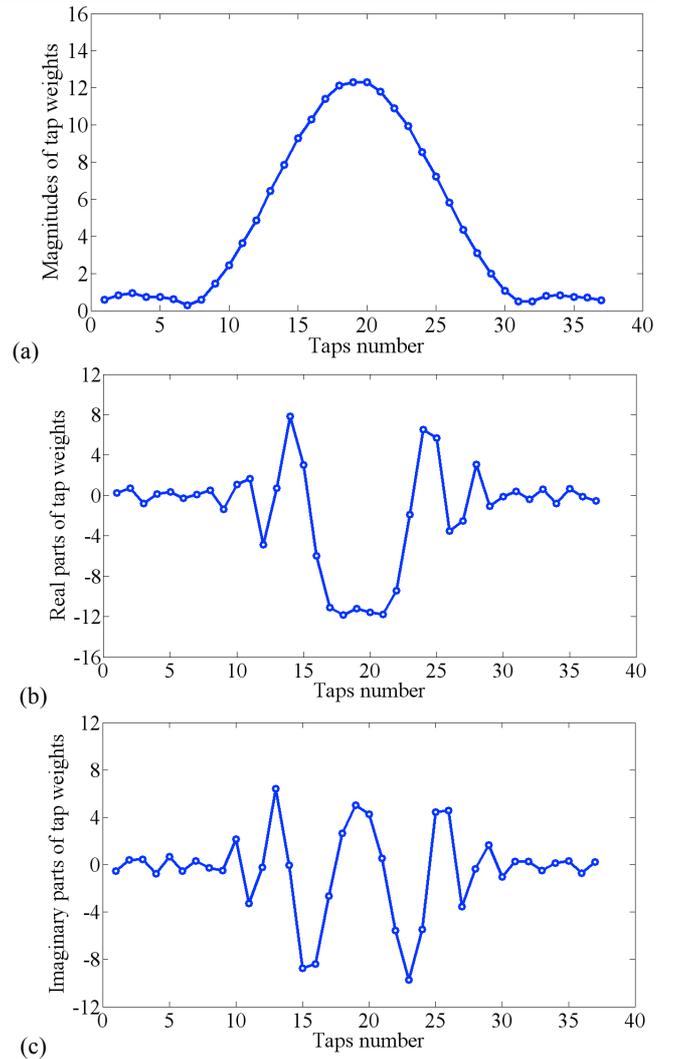

*Fig. 6. Tap weights of variable-step-size LMS adaptive filter (tap orders are centralized), (a) magnitudes of tap weights in VSS-LMS filter, (b) real parts of tap weights in VSS-LMS filter, (c) imaginary parts of tap weights in VSS-LMS filter*

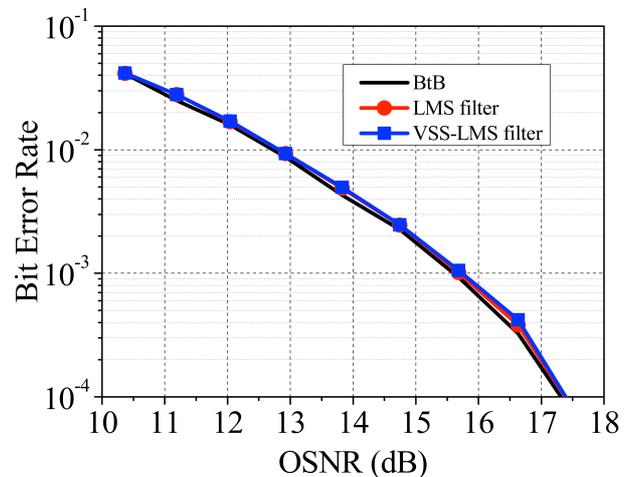

*Fig. 7. CD compensation using traditional LMS and variable-step-size LMS adaptive filters (neglecting fiber loss)*

## 5. Discussions

In above analyses and discussions, only the chromatic dispersion was taken into consideration. Actually, the PMD equalization can also be performed using the LMS algorithm. Thus the combination of CD equalizer and PMD equalizer can be implemented simultaneously using the variable-step-size LMS algorithm.

Meanwhile, the CMA algorithm can also be used for the adaptive chromatic dispersion compensation, while the LMS algorithm is also tolerant to small amounts of laser phase noise [37]. However, for larger phase noise or equalization enhanced phase noise [10, 28], the CMA algorithm is more effective, since the performance of the LMS algorithm will be significantly degraded by the large phase noise. It is also worth noting that both LMS (including VSS-LMS) and CMA algorithms can be applied in communication systems using higher-order modulation formats [26,46-48], where the both approaches can be operated in the decision-direct mode.

In addition, in DSP based coherent communication systems the laser phase noise will interact with the dispersion compensation module to introduce an effect of equalization enhanced phase noise (EEPN) [49,50-52]. In the LMS adaptive dispersion equalization, both transmitter laser phase noise and LO laser phase noise will interact with the dispersion equalization module [53,54], and the system performance is equally influenced by the equalization enhanced transmitter phase noise (EETxPN) and the equalization enhanced LO phase noise (EELOPN).

## 6. Conclusions

A variable-step-size least mean square equalizer has been developed to compensate CD in a 112-Gbit/s PDM-QPSK coherent optical transmission system. The variable-step-size LMS adaptive filter can make a compromise between the algorithm convergence speed and the CD compensation performance compared with traditional LMS adaptive filter. The tap weights in the traditional LMS filter and the VSS-LMS filter have been analyzed, and the CD compensation effects using the two adaptive filters compared by evaluating the BER versus OSNR behavior using numerical simulations. It was found that the variable-step-size LMS equalizer can achieve the same performance with a lower complexity, compared to the traditional LMS algorithm.

### Acknowledgements

This work is supported by UK EPSRC project UNLOC EP/J017582/1, EU project GRIFFON 324391, EU project ICONE 608099, and Swedish Vetenskapsradet 0379801.

[*]Corresponding author: tianhua.xu@warwick.ac.uk